\documentclass{article}
\usepackage{arxiv}

\usepackage[utf8]{inputenc}
\usepackage[T1]{fontenc}
\usepackage{amsmath,amssymb,amsfonts}
\usepackage{algorithmic}
\usepackage{graphicx}
\usepackage{textcomp}
\usepackage{xcolor}
\usepackage{booktabs}
\usepackage{multirow}
\usepackage{array}
\usepackage{xurl}
\usepackage{pgfplots}
\usepackage{pgfplotstable}
\usepackage{tikz}
\usepackage{balance}
\usepackage{microtype}
\usepackage[numbers,sort&compress]{natbib}
\usepackage[hidelinks]{hyperref}
\usepackage{doi}

\pgfplotsset{compat=1.18}
\usetikzlibrary{patterns,shapes,arrows,positioning,fit,calc}

\def\BibTeX{{\rm B\kern-.05em{\sc i\kern-.025em b}\kern-.08em
    T\kern-.1667em\lower.7ex\hbox{E}\kern-.125emX}}

\begin{document}

\title{OrbitAI : Evaluating Tool-Augmented Agents for Enterprise Question Answering}

\author{
Rishabh Gupta, Uday Kharatmol, Krishna Bhatt,\\
Avadhesh Nagar, Naman Sanotiya, and Tarun Jain\\
\textit{Department of Computer Science and Engineering,}\\
\textit{Indian Institute of Information Technology Vadodara, Gandhinagar, Gujarat, India}\\
\texttt{\small rishabh\_gupta@iiitvadodara.ac.in}, \texttt{\small 20261603031@iiitvadodara.ac.in}, \texttt{\small 20261603014@iiitvadodara.ac.in}\\
\texttt{\small 202451185@iiitvadodara.ac.in}, \texttt{\small 202451114@iiitvadodara.ac.in}, \texttt{\small 202451160@iiitvadodara.ac.in}
}
\date{}
\maketitle

\begin{abstract}
Enterprise information is commonly stored across multiple platforms rather than in one centralized location. Answering questions about workplace activities may require information from several sources, including project documents, task management systems, and team conversations. Conventional AI-based question-answering systems mainly focus on retrieving  information from a single source, which also limits their ability to answer questions that depend on multiple sources. Therefore, we evaluate Orbit AI, a tool-based retrieval-augmented generation (RAG) system that uses a ReAct agent implemented with LangGraph to retrieve and combine information from three different workspace sources. The evaluation uses a deterministic benchmark of 100 questions based on a synthetic enterprise workspace. We compare three configurations: (1) a standalone LLM without retrieval, (2) document-based RAG, and (3) a multi-tool agent with access to all three sources..

The held out evaluation set contains  80 questions designed for Phase~3 workspace; Gemini~3.5~Flash~Lite achieves an accuracy of 86.25\% with complete agent. Document only RAG achieves 60.00~\% (26.25~pp higher), and the plain LLM achieves 23.13~\%. This gap is even larger for multi source questions, as the full agent answers 84.38\%, while document only RAG answers 40.63\%, a 43.75~pp increase. This is a successful outcome of the retrieval from all three workspace choices, where an answer is dependent on information spread over all three choices.

The evaluated Llama~3.2 configuration with local inference achieves 5.50\% overall correctness under rubric v2.2. The Llama~3.2 full agent configuration achieves a held out accuracy of 52.50\%, which is 1.88~pp higher than the accuracy of document only RAG. The gain of Gemini is equivalent to 26.25~pp. The comparison must be taken with a pinch of salt, because Gemini was accessed via an API hosted by them, while Llama was run locally via Ollama and the models are of different size. It is therefore not possible to isolate out the cause of the performance gaps in this experiment.

We present results for both evaluation and report on two full-agent responses from Gemini which were scored post-hoc. The paper also addresses the constraints of the evaluation such as the small artificial workspace and the similarity of the Q53 problem to a problem in the system prompt.
\end{abstract}

\keywords{retrieval augmented generation \and tool augmented agents \and enterprise AI \and multi source question answering \and LangGraph \and evaluation methodology \and workspace intelligence}

\section{Introduction}
\label{sec:intro}

Software teams typically has several systems that provide information. Specifications, policies and plans contain the long form of a project's knowledge, task trackers, due dates and statuses, and chat records and updates. To answer a question like ``Who owns the document export feature, and what is it's due date?'' you need at least two of these sources. The specification is the owner and the task tracker has the due date. Both sources are not the only answers.

Retrieval augmented generation (RAG) has consistently performed on the baselines of large language models (LLMs) on document question answering tasks~\cite{lewis2020rag,gao2023survey}. Conventional RAG pipeline, on the other hand, fetches data from a single vector. Task records and chat history cannot be available without them being added to that vector index. The one possibility is to merge all the sources into a single unique index, which may not reveal the original data structure. The other one is to trigger the specific tools at inference time~\cite{schick2023toolformer,yao2023react}.

Orbit AI’s tool-based method for extracting and integrating information from disparate sources in an enterprise workspace. Many studies have been conducted on the use of machine learning and intelligent computing for cloud security, privacy preservation, malicious-user prediction, and threat detection \cite{qmlmalicious2022,gupta2020mlpam,saxena2023vmthreat,privacyDifferential2022,iotprotection2022,differentialClassification2022,medicalPrivacy2022,deepPrivacy2023,accessPolicy2022,privacyCloud2022,attributeEncryption2020,outsourcedPrivacy2020,fogPrivacy2022,differentialSharing2022,cryptographyStorage2023,differentialPrivacy2023,outsourcedData2022}. Related contributions have also investigated federated learning, multi-cloud architectures, secure resource management, and intelligent applications in smart grids, healthcare and speech processing \cite{fedmup2024,saxena2021multicloud,vmworkload2023,maids2024,clientDetection2022,smartgridforecast2023,emotionReview2020,hedcm2021,mentalHealthEnsemble2022,vehicleRouting2025}. Collectively, these studies demonstrate the wider application of machine learning and intelligent computing to build secure, distributed and data-driven systems.

Recently, intelligent cloud environments, adaptive cybersecurity frameworks, federated learning, and AI-driven threat intelligence have gained increasing attention to enhance the security of distributed computing infrastructures \cite{shamAI2026,collaborativeCloudIoT2026,htarfNet2026,globalCloudCyber2026,greenCloudIntelligence2026,multifactorTrust2026,adaptiveCyberThreat2026,intelligentCyberAnalytics2025,metaUnifiedCyber2025,selfAdaptiveCyber2025,secureCloudToffoli2025}. Complementary studies have investigated large language model ecosystems, heterogeneous data management, cloud workload optimization, resource management, and foundation models for computing systems \cite{llmSupplyChain2025,quadCaching2025,reeTM2025,adaptiveQuantumCloud2025,hybridPSOCloud2025,selfHealingDigitalTwin2025,secureResourceManagement2025,multiVMForecast2025,foundationOS2023}. Other contributions have investigated energy-efficient virtual machine placement \cite{hybridEnergyVMP2026}, neural network-based resource reservation \cite{quantumBlackholeCloud2025}, workload pattern analysis and dynamic load balancing in cloud data centers \cite{adaptiveEvolutionaryCloud2025,workloadPatternCloud2025,dynamicCloudLoadBalancer2025}. Research has also explored temporal pattern discovery, predictive load forecasting, resource provisioning and dynamic caching in heterogeneous data environments \cite{fuzzyPeriodicPatterns2025,evolutionarySmartGrid2025,oversubscriptionCloud2024,latencyCachingDatalake2024,secureProvisioningCloud2024}. Taken together the papers demonstrate the increasing application of intelligent and adaptive methods in the fields of cloud computing, cybersecurity, data management and resource optimization. Given these developments in intelligent and distributed systems, we present in this paper Orbit AI, a multi-tool question answering system for retrieving and combining information from multiple sources in an enterprise workspace. LangGraph~\cite{langgraph2024} was used to implement the system, which is a ReAct-based agent with access to three tools, \texttt{search\_workspace\_tool} for vector-similarity search over embedded documents, \texttt{get\_tasks\_tool} for SQL-based task lookup, and \texttt{get\_recent\_chat\_tool} for retrieving recent chat messages. We evaluate the system using the hosted API of Google Gemini~3.5~Flash~Lite~\cite{gemini2024} and the local Ollama app for Meta Llama~3.2 (3.2B parameters)~\cite{llama2024}. We evaluate each model on three settings: a standalone LLM, document-only retrieval augmented generation (RAG), and a full multi-tool agent. This evaluation examines the effect of having access to multiple sources of information and tools on the performance of answering questions, in comparison to document-based and standalone approaches.

\textbf{Research Questions.}
\begin{enumerate}
    \item Does a full multi tool agent access improves factual correctness over document only RAG on a workspace benchmark, particularly for multi source questions?
    \item Does a smaller LLM derive same benefit from tool augmentation as a larger hosted model under the same benchmark?
    \item What are the primary failure modes of each configuration?
\end{enumerate}

\textbf{Primary Contribution.}
This work provides a deterministic benchmark consisting of 100 questions, which are assessed in a synthetic Phase~3 workspace with 3 documents, 16 tasks, and 16 chat messages. Additionally, a post-hoc v2.2 correction changes two Gemini records. For the held out multisource subset, the accuracy of Gemini increases by 43.75~pp when using the tools as opposed to document only RAG. The improvement for the evaluated Llama~3.2 configuration is much smaller, suggesting that the advantage of tool access is different between the two models tested.

\section{Background and Related Work}
\label{sec:related}

\subsection{Retrieval-Augmented Generation}
Retrieved passages are used to bound the responses in RAG~\cite{lewis2020rag} for the purpose to minimize hallucinations on knowledge intensive tasks. In production, embedding similarity~\cite{karpukhin2020dpr} is widely used. In recent surveys~\cite{gao2023survey}, RAG methods are categorized based on various factors, including retrieval granularity, query routing, and result fusion. In this study, we use the \texttt{rag\_docs\_only} configuration, which fetches the top-2 chunks from a pgvector store~\cite{pgvector2024} and appends them to the system prompt.

\subsection{Labeling Tool Augmented and Agentic LLMs}
The model, Toolformer~\cite{schick2023toolformer} showed that it can learn to use external APIs when generating. ReAct~\cite{yao2023react} introduces the combination of reasoning and using tools in an iterative manner and LangGraph~\cite{langgraph2024} introduces a statefull agent graph API for runtime whose agents can revisit previous steps. The \texttt{full\_agent} setup uses a LangGraph based ReAct agent that can access all the three workspace sources. Also the tools it uses depends on the user's query.

\subsection{Enterprise Question Answering}
Enterprise question answering involves challenges which are not commonly found in the open domain QA.Information may be spread across different sources, and the system must respect the privacy and restrictions while avoiding the answer that cannot be supported by available evidence in the knowledge base 
~\cite{asai2023selfrag,shi2023trustworthy}. In conventional RAG system , retrieval mainly focus on documents , leaving structured information from project management tool , and calendars which are outside the usual retrieval process.Orbit addresses this limitation by providing separate tools for accessing these sources, allowing the agent to retrieve the information needed to answer a query. 

\subsection{Deterministic Evaluation}
Although flexible, LLM-as-judge evaluation~\cite{zheng2023llm-judge} can be hard to interpret and can be prone to some bias. Instead, a rubric of a deterministic nature using a keyword match approach to evaluation (Section~\ref{sec:eval_method}) is used. This decision constrains the range of semantically similar answers the scorer can identify, simplifies the scoring process and also enhances the process~\cite{muhlgay2023generating}

\section{System Architecture}
\label{sec:architecture}

\begin{figure}[htbp]
\centering
\resizebox{\columnwidth}{!}{%
\begin{tikzpicture}[
  box/.style={
    draw, rounded corners=3pt,
    minimum width=2.35cm,
    minimum height=0.65cm,
    font=\footnotesize,
    align=center
  },
  tool/.style={
    draw, rounded corners=3pt,
    minimum width=1.85cm,
    minimum height=0.48cm,
    font=\scriptsize,
    align=center,
    fill=gray!15
  },
  arr/.style={->, >=stealth, thick}
]

\node[box, fill=blue!10] (user) at (0,5.0)
    {User Query};

\node[box, fill=orange!10] (api) at (0,3.9)
    {FastAPI Backend};

\node[box, fill=yellow!10] (router) at (0,2.8)
    {Configuration Router};

\node[box, fill=red!8] (plain) at (-2.6,1.5)
    {Plain LLM\\{\scriptsize (no retrieval)}};

\node[box, fill=green!8] (rag) at (0,1.5)
    {RAG\\{\scriptsize (docs only)}};

\node[box, fill=purple!8] (agent) at (2.6,1.5)
    {LangGraph Agent\\{\scriptsize (multi-tool ReAct)}};

\node[tool] (vec) at (0,0.25)
    {pgvector-chunks};

\node[tool] (t1) at (2.6,0.1) {search\_docs};
\node[tool] (t2) at (2.6,-0.55) {get\_tasks};
\node[tool] (t3) at (2.6,-1.2) {get\_chat};

\draw[dashed, rounded corners=4pt]
    (1.55,0.48) rectangle (3.65,-1.55);

\node[font=\scriptsize] at (2.6,0.62) {Available Tools};

\node[box, fill=blue!15, minimum width=6.5cm]
    (llm) at (0,-2.15)
    {LLMs};

\coordinate (llm_plain) at (-2.6,-1.825);
\coordinate (llm_rag) at (0,-1.825);
\coordinate (llm_agent) at (2.6,-1.825);

\node[box, fill=gray!15] (resp) at (0,-3.2)
    {Evidence};

\draw[arr] (user) -- (api);
\draw[arr] (api) -- (router);

\draw[arr] (router.west) -| (plain.north);
\draw[arr] (router.south) -- (rag.north);
\draw[arr] (router.east) -| (agent.north);

\draw[arr] (plain.south) -- (llm_plain);

\draw[arr] (rag.south) -- (vec.north);
\draw[arr] (vec.south) -- (llm_rag);

\draw[arr] (agent.south) -- (2.6,0.85);

\draw[arr] (2.6,-1.55) -- (llm_agent);

\draw[arr] (llm.south) -- (resp.north);

\end{tikzpicture}%
}
\caption{Orbit AI system architecture.}
\label{fig:architecture}
\end{figure}
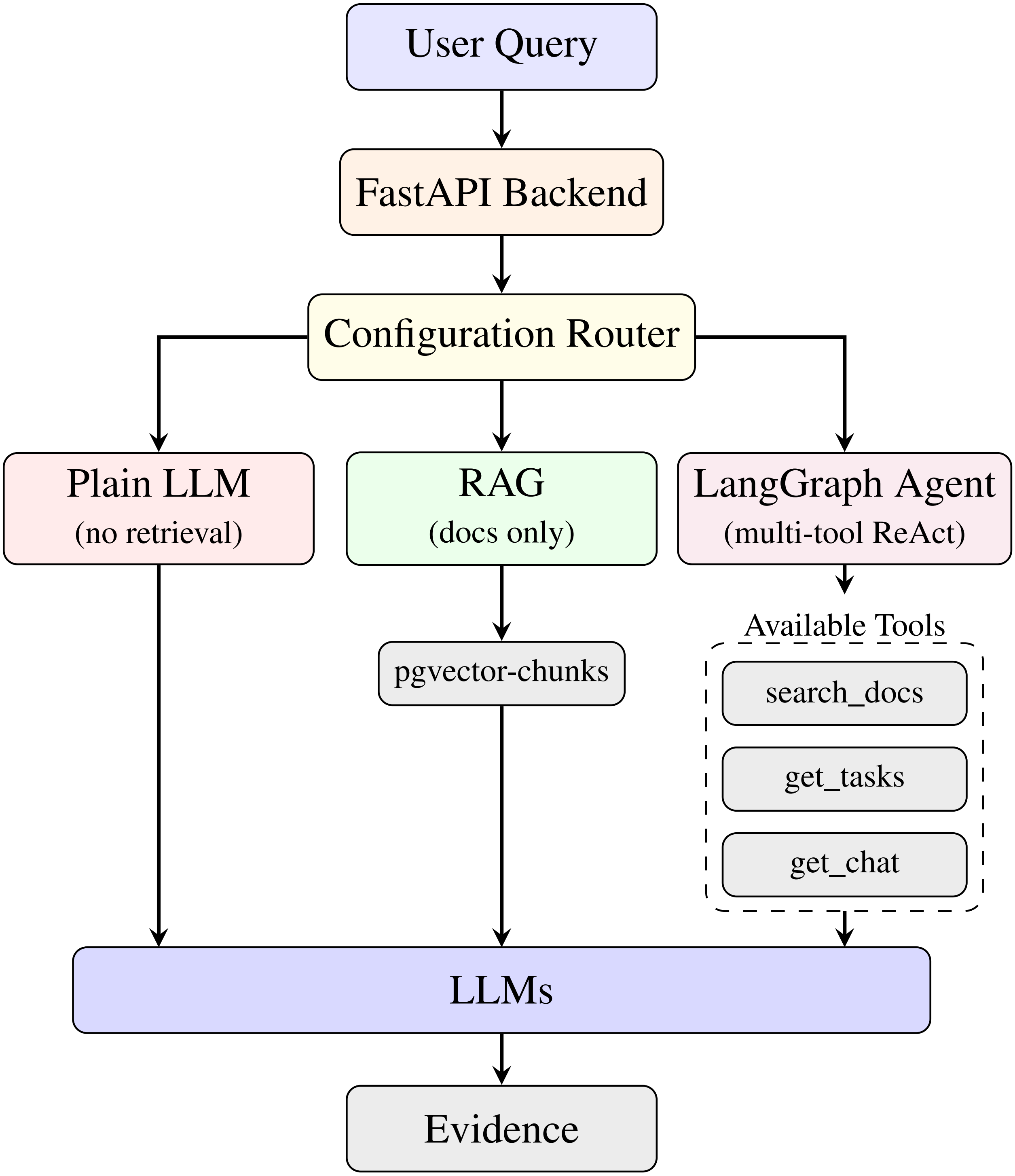

Figure~\ref{fig:architecture} shows the main request flow in Orbit AI. The FastAPI backend receives a user query and directs it to one of the three evaluation configurations. In the full agent configuration, a LangGraph ReAct loop can call the three workspace tools. Whenever a configuration uses document retrieval, the top-2 retrieved chunks are inserted before model execution.

\subsection{Document Chunking}
The chunker (\texttt{backend/app/rag/chunker.py}) chunks the documents in a window of \textbf{1,000 words} and an overlap of \textbf{50 words}. Each of the three Phase~3 documents has only a few words (111, 108, and 120) and thus represents a single chunk. As a result, the benchmark provides limited evidence of retrieval of multiple chunks (or multiple larger documents) as well as of any precision over a large document collection. In the case of document enabled configurations, vector search will take the most similar top-2 chunks and add them to the system prompt.

\subsection{Workspace Scope}
The Phase~3 synthetic workspace (Workspace~3, UUID \texttt{00000000-0000-4000-8000-000000000003}) contains
\begin{itemize}
    \item 3 documents: Architecture Specification, Security Policy, Release Plan Sprint~15
    \item 16 tasks (T01--T16) with assignees, statuses, and due dates
    \item 16 chat messages covering team coordination and status updates
\end{itemize}

\section{Benchmark Dataset}
\label{sec:dataset}

\subsection{Dataset Composition}

The 100 benchmark questions are divided into two cohorts and assigned to one of three semantic categories, as summarized in Table~\ref{tab:dataset}.

\begin{table}[htbp]
\caption{Dataset Composition by Cohort and Category}
\label{tab:dataset}
\centering
\resizebox{\columnwidth}{!}{%
\begin{tabular}{lcccc}
\toprule
\textbf{Cohort} & \textbf{Single Source} & \textbf{Multi Source} & \textbf{Unanswerable} & \textbf{Total} \\
\midrule
Baseline (Q01--Q20) & 8 & 8 & 4 & 20 \\
Held-Out (Q21--Q100) & 32 & 32 & 16 & 80 \\
\midrule
\textbf{Full Dataset (Q01--Q100)} & \textbf{40} & \textbf{40} & \textbf{20} & \textbf{100} \\
\bottomrule
\end{tabular}%
}
\end{table}

\textbf{Category definitions}:
\begin{itemize}
    \item \textbf{Single source:} Answer can be found using information from just one workspace such as documents.
    \item \textbf{Multi source:} Answering the questions requires information from at least two different  workspace sources.
    \item \textbf{Unanswerable:} The workspace does not contain the information needed to answer the question.In such cases , the system is expected to refuse rather than provide an unsupported answer.
\end{itemize}

\subsection{Cohort Interpretation}
\label{sec:cohorts}
The \textbf{Held Out Cohort} (Q21--Q100, $n=80$) serves as the primary Phase~3 evaluation set. All 80 questions were prepared specifically for Workspace~3 and were separated from the initial development of the scoring rules. However, two Gemini responses (Q93 and Q99) from this cohort were re-evaluated following the v2.2 scoring rule correction, as disclosed in Section~\ref{sec:limitations}. Therefore, the held-out cohort should not be considered completely independent of the scoring process.

The \textbf{Baseline Cohort} (Q01--Q20, $n=20$), on the other hand , was retained for diagnostic purposes. These questions came from earlier evaluation phases. Half of them refer to entities , including the Sprint 14 Redis tasks and OAuth workflows , that are no longer present in Workspaces 3. However , the entire cohort cannot be treated unanswerable . For example, Q01 can still be answered using the current Security Policy. While several other questions concern information that remains available in the Phase 3 documents.

\textbf{Cohort distinction:} The two cohorts are assigned to different relationship to the workspace in Phase~3. If the answer to some of the baseline questions is an abstention, it does not mean that the question itself is unanswerable, it merely means that the answer is \textbf{not available} at the moment: the needed information is \emph{missing from the current workspace}. The two groups combines into a single average would mask this difference. We therefore consider the results of the tests held out as the most important evidence to be reported and baseline results are reported separately for diagnostic purposes.

\subsection{Scoring Methodology}
\label{sec:eval_method}

A deterministic, keyword and regex based scoring system (Rubric~v2.2 deterministic audited) is used. In the case of an answerable question, the scorer verifies the expected target values, the required concepts and absence of known forbidden values. If there are no contradictions, at least half of the required concepts are matched and all target-value groups are matched, a response earns 1.0. A response is awarded 0.5 if it matches at least half of the groups of values in the target, but not all. Also included are responses that match at least half of the groups of values in the target with only one contradiction in a partial match. A response will earn 0.0 if it does not contain the required information or if it contains information that contradicts the rubric's conditions of incorrectness. If no target value has been given for the question, the scorer has as a last resort, the token overlap with the reference answer.

If a question is not answerable, a response is considered to be correct with 1.0 if it contains a refusal signal or a required refusal concept, but does not assert a forbidden fabricated value. It is set to 0.0 if it does not accept that it is missing information, or claims an impossible made-up number. The scorer is not a full-fledged semantic evaluator; it relies on pre-determined lexical patterns, synonym normalization and contradiction checks.

\textbf{Faithfulness} indicates whether the answer is based on the information for the model. Different configurations have different evaluation populations. In the case of \texttt{plain\_llm}, faithfulness is computed using the 20 questions that do not have a correct answer, as there is no workspace retrieval evidence in this setting. For \texttt{rag\_docs\_only} and \texttt{full\_agent}, faithfulness is measured on all 100 questions, based on the evidence that is retrieved or provided to the model. When a question cannot be answered, a response is faithful if it admits that there is not enough evidence in the workspace to answer the question, but does not include any forbidden fabricated facts. When answers are required, the assessment determines if the target facts are supported by the evidence retrieved. Responses with no evaluable evidence or with execution errors are not included in the average for faithfulness. The scores for different configurations therefore cannot be directly compared and should not be interpreted as the same.

\section{Experimental Setup}
\label{sec:setup}
\subsection{Models}
\begin{itemize}
    \item \textbf{Gemini~3.5~Flash~Lite [6]:} Evaluated using the hosted Google Gemini API.
    \item \textbf{Llama~3.2 (3.2B) [7]:} Evaluated locally using Ollama.
\end{itemize}

\subsection{Configurations}
The three configurations differ in the resources made available to the model:
\begin{itemize}
    \item \textbf{plain\_llm:} The model has no access to workspace retrieval tools and generates responses using only its pretrained knowledge.
    \item \textbf{rag\_docs\_only:} Similarity search retrieves the top two document chunks and adds them to the model context. The model has no access to task records or chat history.
    \item \textbf{full\_agent:} The model receives the same initial document context and can use a LangGraph ReAct agent to retrieve additional information from documents, task records, and recent chat messages.
\end{itemize}

\subsection{Infrastructure}
\begin{itemize}
    \item \textbf{Backend:} FastAPI (Python), PostgreSQL, and Supabase with pgvector.
    \item \textbf{Gemini inference:} Hosted Google Gemini API.
    \item \textbf{Llama inference:} Local Ollama deployment.
    \item \textbf{Benchmark size:} 100 questions evaluated across three configurations for each model, resulting in 300 evaluation attempts per model.
    \item \textbf{Experimental repetitions:} Each question/configuration combination was evaluated once.
\end{itemize}

\subsection{System Prompt Proximity}
\label{sec:q53_note}
The complete agent system prompt includes an example of when the agent should invoke \texttt{get\_tasks\_tool}: " Priya's document export task due date ". The task deadline specified in Q53 refers to the deadline of the document export feature and is assigned to Priya. The similarity between the sample and the question may lead to a possible prompt priming effect. We did not test a modified version of the original prompt and this is therefore treated as a limitation in Section~\ref{sec:limitations}. The answer to Q53 also depends on the release-candidate date which is therefore needed in order to answer the question.

\section{Results}
\label{sec:results}

The results below use rubric v2.2-deterministic-audited for Llama and the corrected v2.2 rubric for Gemini.

\subsection{Overall Correctness}

Table~\ref{tab:overall} shows mean correctness for the complete benchmark (Q01--Q100) and for the held-out cohort alone (Q21--Q100).

\begin{table}[htbp]
\caption{Overall Correctness by Model and Configuration}
\label{tab:overall}
\centering
\resizebox{\columnwidth}{!}{%
\begin{tabular}{llcc}
\toprule
\textbf{Model} & \textbf{Configuration} & \textbf{Full (n=100)} & \textbf{Held-Out (n=80)} \\
\midrule
Gemini 3.5 FL         & plain\_llm      & 0.2250 & 0.2313 \\
                      & rag\_docs\_only  & 0.5300 & 0.6000 \\
                      & full\_agent     & \textbf{0.7450} & \textbf{0.8625} \\
\midrule
Llama 3.2             & plain\_llm      & 0.0550 & 0.0500 \\
Llama 3.2             & rag\_docs\_only  & 0.4550 & 0.5063 \\
                      & full\_agent     & 0.4700 & 0.5250 \\
\bottomrule
\multicolumn{4}{l}{\footnotesize Gemini: rubric v2.2-corrected. Llama: rubric v2.2-deterministic-audited.}\\
\end{tabular}%
}
\end{table}

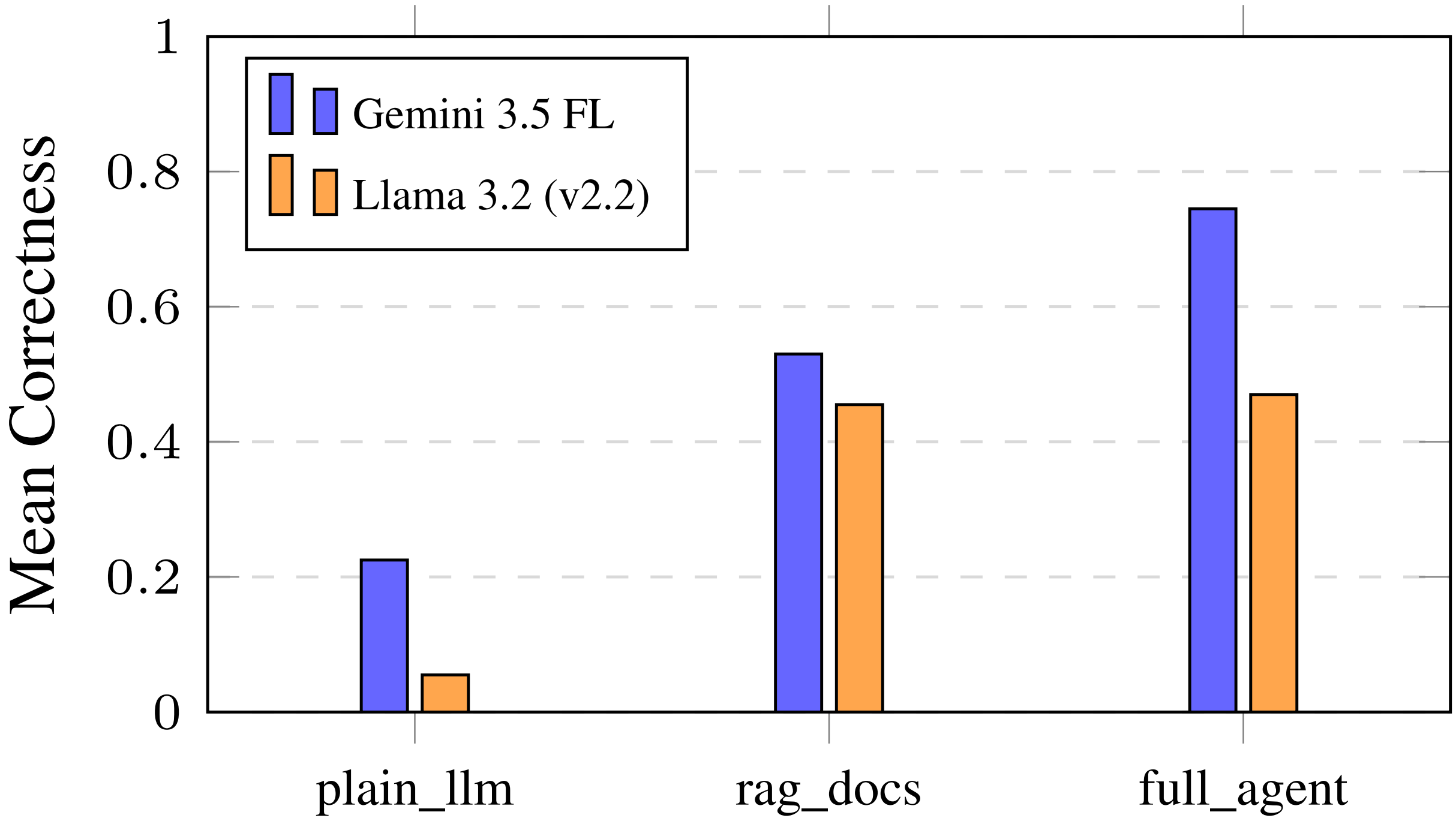
\begin{figure}[htbp]
\centering
\resizebox{0.98\columnwidth}{!}{%
\begin{tikzpicture}
\begin{axis}[
    ybar,
    bar width=0.22cm,
    width=7.5cm,
    height=4.8cm,
    ylabel={Mean Correctness},
    ylabel style={font=\footnotesize},
    ymin=0, ymax=1.0,
    ytick={0,0.2,0.4,0.6,0.8,1.0},
    yticklabel style={font=\scriptsize},
    xtick=data,
    xticklabels={plain\_llm, rag\_docs, full\_agent},
    xticklabel style={font=\scriptsize},
    legend style={at={(0.03,0.97)}, anchor=north west, font=\tiny, legend cell align=left},
    enlarge x limits=0.25,
    ymajorgrids=true,
    grid style={dashed, gray!30},
]
\addplot[fill=blue!60] coordinates {(1,0.2250)(2,0.5300)(3,0.7450)};
\addplot[fill=orange!70] coordinates {(1,0.0550)(2,0.4550)(3,0.4700)};
\legend{Gemini 3.5 FL, Llama 3.2 (v2.2)}
\end{axis}
\end{tikzpicture}%
}
\caption{Overall correctness on the full 100-question dataset.}
\label{fig:overall_correctness}
\end{figure}

Figure~\ref{fig:overall_correctness} compares overall correctness across configurations. The figure reports the archived rubric v2.2-deterministic-audited scores for all three Llama configurations.

\subsection{Category Level Results}

Table~\ref{tab:category} breaks down correctness of questions by category across the all 100 questions. The denominators used for faithfulness on unanswerable questions are reported separately in Table~\ref{tab:faithfulness}.

\begin{table}[htbp]
\caption{Category-Level Mean Correctness (Full Dataset, N=100)}
\label{tab:category}
\centering
\resizebox{\columnwidth}{!}{%
\begin{tabular}{llccc}
\toprule
\textbf{Model} & \textbf{Configuration} & \textbf{Single (n=40)} & \textbf{Multi (n=40)} & \textbf{Unans. (n=20)} \\
\midrule
Gemini 3.5 FL & plain\_llm      & 0.0625 & 0.0000 & 1.0000 \\
              & rag\_docs\_only  & 0.4875 & 0.3375 & 1.0000 \\
              & full\_agent     & \textbf{0.6625} & \textbf{0.7000} & 1.0000 \\
\midrule
Llama 3.2     & plain\_llm      & 0.0750 & 0.0125 & 0.1000 \\
              & rag\_docs\_only  & 0.4500 & 0.3125 & 0.7500 \\
              & full\_agent     & 0.4250 & 0.4000 & 0.7000 \\
\bottomrule
\multicolumn{5}{l}{\footnotesize Llama values use rubric v2.2-deterministic-audited.} \\
\end{tabular}%
}
\end{table}

\textbf{Key observations:}
\begin{itemize}
    \item Gemini achieves 100\% on all unanswerable questions across all configurations under rubric v2.2-corrected.
    \item Under rubric v2.2, Llama scores 10\% on plain\_llm unanswerable questions, 75\% on rag\_docs\_only, and 70\% on full\_agent.
    \item Llama's unanswerable score decreases from rag\_docs\_only to full\_agent (0.75 $\to$ 0.70), indicating that tool access does not consistently improve abstention on questions that cannot be answered from the workspace.
\end{itemize}

\subsection{Held-Out Cohort Results}
\label{sec:heldout}

\begin{table}[htbp]
\caption{Phase~3 Held-Out Cohort Results (Q21--Q100, n=80)}
\label{tab:heldout}
\centering
\resizebox{\columnwidth}{!}{%
\begin{tabular}{llccc}
\toprule
\textbf{Category} & \textbf{Configuration} & \textbf{Gemini 3.5 FL} & \textbf{Llama 3.2 (v2.2)} & \textbf{$\Delta$ (G$-$L)} \\
\midrule
\multirow{3}{*}{Single (n=32)} & plain\_llm      & 0.0781 & 0.0781 & 0.00 pp \\
                               & rag\_docs\_only  & 0.5938 & 0.5313 & +6.25 pp \\
                               & full\_agent     & \textbf{0.8125} & 0.5000 & +31.25 pp \\
\midrule
\multirow{3}{*}{Multi (n=32)}  & plain\_llm      & 0.0000 & 0.0156 & $-$1.56 pp \\
                               & rag\_docs\_only  & 0.4063 & 0.3594 & +4.69 pp \\
                               & full\_agent     & \textbf{0.8438} & \textbf{0.4688} & +37.50 pp \\
\midrule
\multirow{3}{*}{Unans. (n=16)} & plain\_llm      & 1.0000 & 0.0625 & +93.75 pp \\
                               & rag\_docs\_only  & 1.0000 & 0.7500 & +25.00 pp \\
                               & full\_agent     & 1.0000 & 0.6875 & +31.25 pp \\
\midrule
\multirow{3}{*}{Total (n=80)}  & plain\_llm      & 0.2313 & 0.0500 & +18.13 pp \\
                               & rag\_docs\_only  & 0.6000 & 0.5063 & +9.38 pp \\
                               & full\_agent     & \textbf{0.8625} & \textbf{0.5250} & +33.75 pp \\
\bottomrule
\multicolumn{5}{l}{\footnotesize Llama: rubric v2.2-deterministic-audited. Gemini: rubric v2.2-corrected.}\\
\end{tabular}%
}
\end{table}

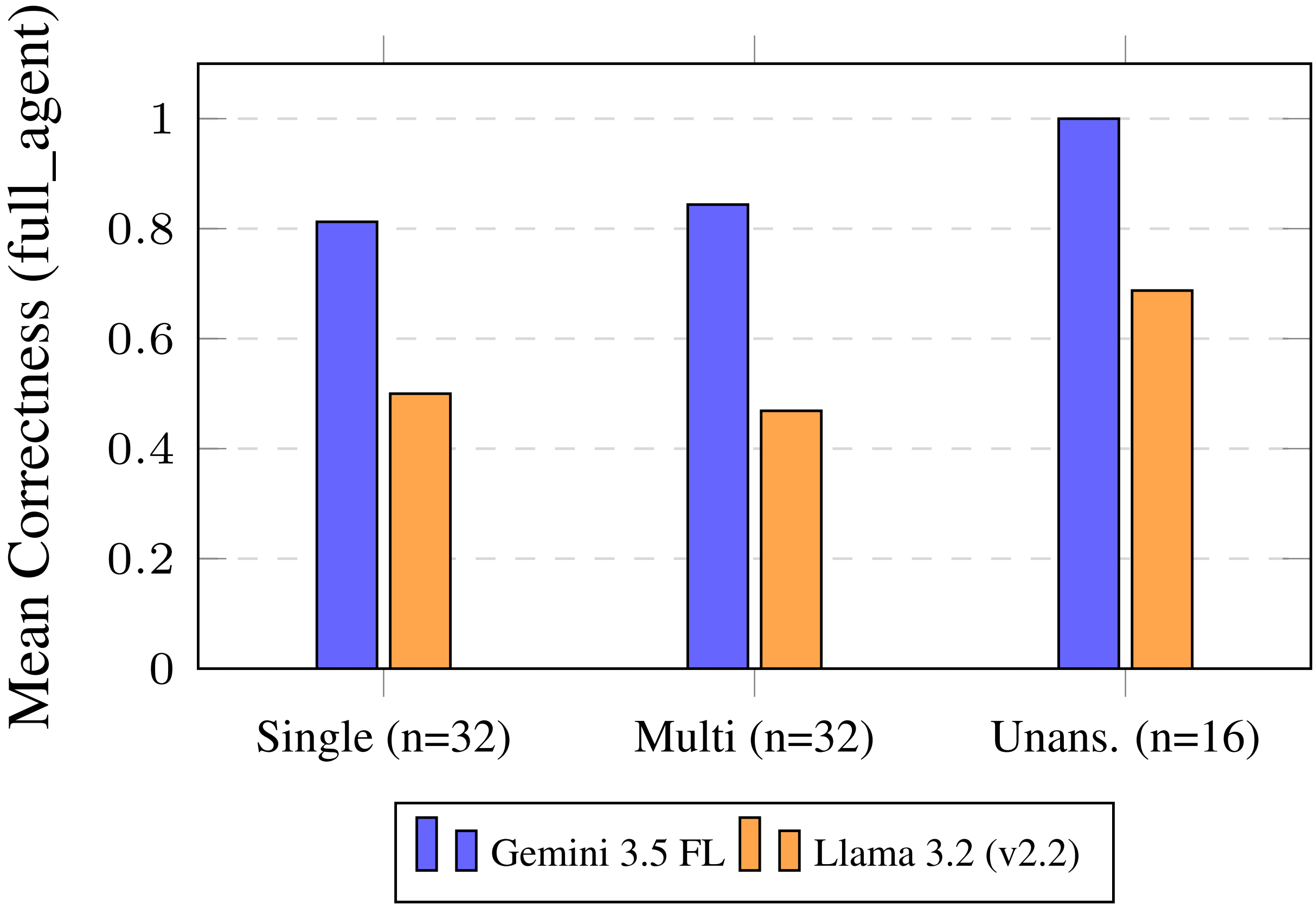
\begin{figure}[htbp]
\centering
\resizebox{0.98\columnwidth}{!}{%
\begin{tikzpicture}
\begin{axis}[
    ybar,
    bar width=0.32cm,
    width=7.5cm,
    height=4.8cm,
    ylabel={Mean Correctness (full\_agent)},
    ylabel style={font=\footnotesize},
    ymin=0, ymax=1.1,
    ytick={0,0.2,0.4,0.6,0.8,1.0},
    yticklabel style={font=\scriptsize},
    xtick=data,
    xticklabels={Single (n=32), Multi (n=32), Unans. (n=16)},
    xticklabel style={font=\scriptsize},
    legend style={at={(0.5,-0.22)}, anchor=north, font=\tiny, legend columns=2},
    enlarge x limits=0.25,
    ymajorgrids=true,
    grid style={dashed, gray!30},
]
\addplot[fill=blue!60] coordinates {(1,0.8125)(2,0.8438)(3,1.0000)};
\addplot[fill=orange!70] coordinates {(1,0.5000)(2,0.4688)(3,0.6875)};
\legend{Gemini 3.5 FL, Llama 3.2 (v2.2)}
\end{axis}
\end{tikzpicture}%
}
\caption{Held-out full agent correctness by question category (Q21--Q100).}
\label{fig:category_correctness}
\end{figure}

Figure~\ref{fig:category_correctness} compares the full agent results across the three question categories.

\subsection{Gemini \& Llama Correctness Gap}

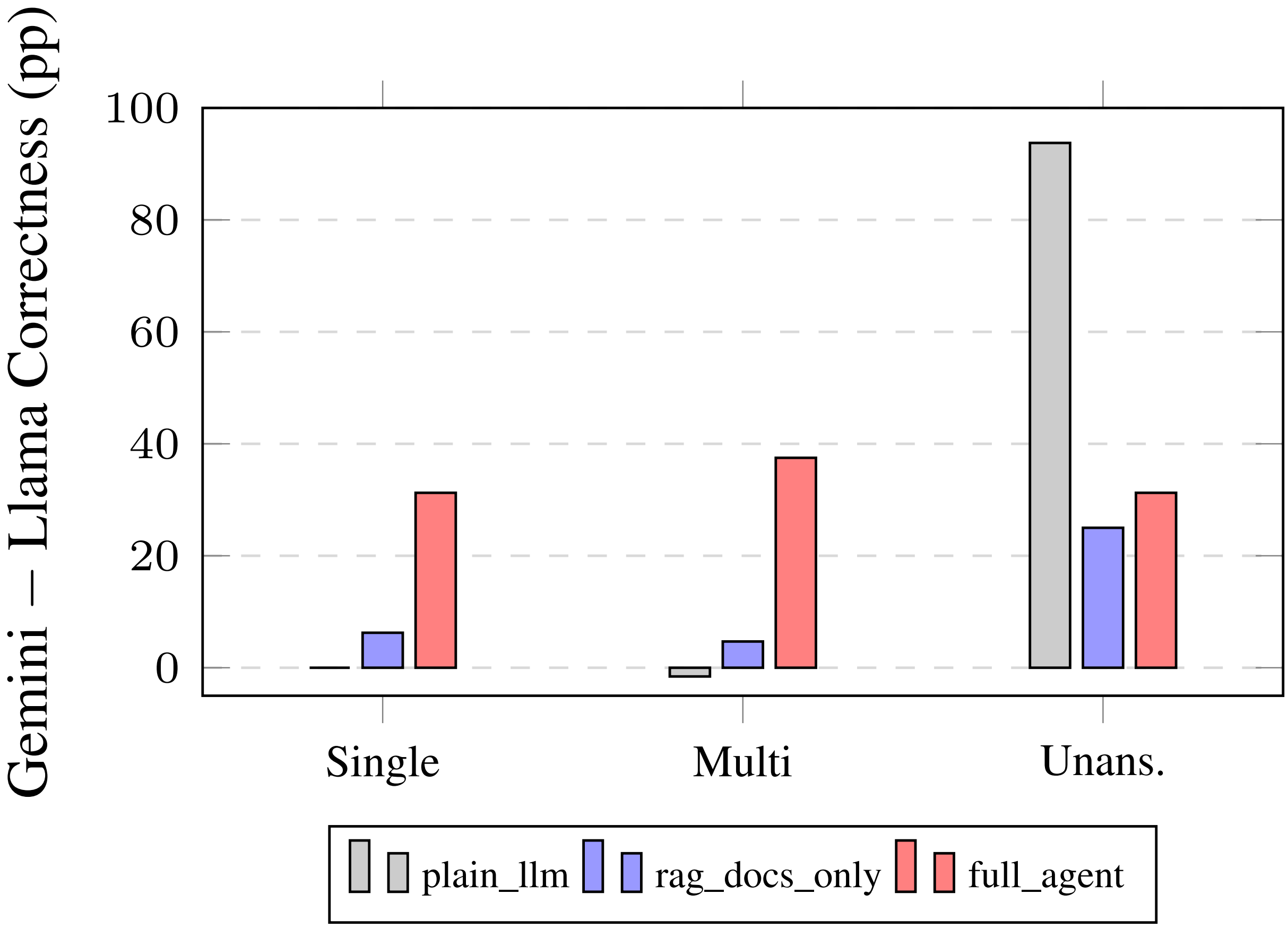
\begin{figure}[htbp]
\centering
\resizebox{0.98\columnwidth}{!}{%
\begin{tikzpicture}
\begin{axis}[
    ybar,
    bar width=0.22cm,
    width=7.5cm,
    height=4.8cm,
    ylabel={Gemini $-$ Llama Correctness (pp)},
    ylabel style={font=\footnotesize},
    ymin=-5, ymax=100,
    ytick={0,20,40,60,80,100},
    yticklabel style={font=\scriptsize},
    xtick=data,
    xticklabels={Single, Multi, Unans.},
    xticklabel style={font=\scriptsize},
    legend style={at={(0.5,-0.22)}, anchor=north, font=\tiny, legend columns=3},
    enlarge x limits=0.25,
    ymajorgrids=true,
    grid style={dashed, gray!30},
]
\addplot[fill=gray!40] coordinates {(1,0.00)(2,-1.56)(3,93.75)};
\addplot[fill=blue!40] coordinates {(1,6.25)(2,4.69)(3,25.00)};
\addplot[fill=red!50] coordinates {(1,31.25)(2,37.50)(3,31.25)};
\legend{plain\_llm, rag\_docs\_only, full\_agent}
\end{axis}
\end{tikzpicture}%
}
\caption{Gemini--Llama correctness gap across held-out question categories (Q21--Q100). Values are Gemini minus Llama percentage points; Llama uses rubric v2.2-deterministic-audited and Gemini uses the corrected v2.2 rubric. Among answerable questions, the gap is largest for full\_agent on single-source (+31.25~pp) and multi-source (+37.50~pp) categories.}
\label{fig:gap}
\end{figure}

Figure~\ref{fig:gap} shows the difference in correctness between Gemini and Llama for each category.

\subsection{Incremental Configuration Gains}

Table~\ref{tab:incremental} summarizes the percentage change associated with each retrieval channel.

\begin{table}[htbp]
\caption{Incremental Correctness Gains (Percentage Points)}
\label{tab:incremental}
\centering
\resizebox{\columnwidth}{!}{%
\begin{tabular}{lrrrr}
\toprule
\textbf{Evaluation Slice} & \multicolumn{2}{c}{\textbf{Gemini 3.5 FL}} & \multicolumn{2}{c}{\textbf{Llama 3.2 (v2.2)}} \\
               & RAG$-$Plain & Agent$-$RAG & RAG$-$Plain & Agent$-$RAG \\
\midrule
Full Dataset (n=100)      & +30.50 & +21.50 & +40.00 & +1.50 \\
Held-Out Total (n=80)     & +36.88 & +26.25 & +45.63 & +1.88 \\
Held-Out Single (n=32)    & +51.56 & +21.88 & +45.31 & $-$3.13 \\
Held-Out Multi (n=32)     & +40.63 & +43.75 & +34.38 & +10.94 \\
Held-Out Unans. (n=16)    &   0.00 &   0.00 & +68.75 & $-$6.25 \\
\midrule
All Single-Source (n=40)  & +42.50 & +17.50 & +37.50 & $-$2.50 \\
All Multi-Source (n=40)   & +33.75 & +36.25 & +30.00 & +8.75 \\
\bottomrule
\multicolumn{5}{l}{\footnotesize Held-out = Q21--Q100. Llama uses rubric v2.2-deterministic-audited.}\\
\end{tabular}%
}
\end{table}

\subsection{Baseline Cohort Results}

Table~\ref{tab:baseline} reports results for the diagnostic cohort (Q01--Q20). Some questions still refer to facts available in the Phase~3 workspace i.e. Q01, for instance, can be answered from the Security Policy. Others refer to entities that are no longer present.

\begin{table}[htbp]
\caption{Baseline Cohort Results (Q01--Q20, n=20; Diagnostic Only)}
\label{tab:baseline}
\centering
\resizebox{\columnwidth}{!}{%
\begin{tabular}{llcc}
\toprule
\textbf{Category} & \textbf{Configuration} & \textbf{Gemini 3.5 FL} & \textbf{Llama 3.2 (v2.2)} \\
\midrule
Single (n=8) & plain\_llm      & 0.0000 & 0.0625 \\
             & rag\_docs\_only  & 0.0625 & 0.1250 \\
             & full\_agent     & 0.0625 & 0.1250 \\
\midrule
Multi (n=8)  & plain\_llm      & 0.0000 & 0.0000 \\
             & rag\_docs\_only  & 0.0625 & 0.1250 \\
             & full\_agent     & 0.1250 & 0.1250 \\
\midrule
Unans. (n=4) & plain\_llm      & 1.0000 & 0.2500 \\
             & rag\_docs\_only  & 1.0000 & 0.7500 \\
             & full\_agent     & 1.0000 & 0.7500 \\
\midrule
Total (n=20) & plain\_llm      & 0.2000 & 0.0750\\
             & rag\_docs\_only  & 0.2500 & 0.2500 \\
             & full\_agent     & 0.2750 & 0.2500 \\
\bottomrule
\multicolumn{4}{l}{\footnotesize Llama values use rubric v2.2-deterministic-audited.}\\
\end{tabular}%
}
\end{table}

\subsection{Faithfulness}
\label{sec:faithfulness}
Correctness and faithfulness measure different aspects of model performance. \textbf{Correctness} measures whether the response contains the expected benchmark answer according to the deterministic scoring rubric. \textbf{Faithfulness} measures whether the factual claims made by the model are supported by the evidence actually available to it. A response may be faithful but incorrect if it provides only partially supported information or appropriately acknowledges missing evidence without answering an answerable question. Conversely, a response may contain expected answer values while also introducing unsupported claims.

The faithfulness evaluation is a deterministic evidence-consistency proxy based on retrieved evidence, expected target values, and predefined forbidden values. It is not a comprehensive semantic assessment of every factual statement in a response. Furthermore, the number and type of responses included in the faithfulness calculation vary by configuration. As a result, the values in Table~\ref{tab:faithfulness} should not be compared directly across all configurations.

\begin{table}[htbp]
\caption{Faithfulness by Model and Configuration (Clarified Denominators)}
\label{tab:faithfulness}
\centering
\resizebox{\columnwidth}{!}{%
\begin{tabular}{llccc}
\toprule
\textbf{Model} & \textbf{Configuration} & \textbf{Mean Faith.} & \textbf{N eval.} & \textbf{Evaluated Population} \\
\midrule
Gemini 3.5 FL & plain\_llm      & 1.0000 & 20  & Unanswerable only \\
              & rag\_docs\_only  & 0.8350 & 100 & All questions \\
              & full\_agent     & 0.9000 & 100 & All questions \\
\midrule
Llama 3.2        & plain\_llm   & 0.1000 & 20  & Unanswerable only \\
Llama 3.2        & rag\_docs\_only & 0.7550 & 100 & All questions \\
                 & full\_agent    & 0.7500 & 100 & All questions \\
\bottomrule
\multicolumn{5}{l}{\footnotesize Gemini v2.2-corrected; Llama v2.2-deterministic-audited.}\\
\multicolumn{5}{l}{\footnotesize \textbf{Warning:} Do not compare plain\_llm (n=20) directly with RAG/agent (n=100).}\\
\end{tabular}%
}
\end{table}

\textbf{Key faithfulness findings:}
\begin{itemize}
    \item Under rubric v2.2-deterministic-audited, Llama plain\_llm has a faithfulness score of 0.10 on unanswerable questions.
    \item The 20 abstractive plain based LLM responses from Gemini have a plain\_llm faithfulness score of 1.00 because they all were get from values that are forbidden and did not fabricate any.
    \item The two \textbf{1.00} measurements above are not \textbf{comparable} with RAG agent faithfulness. It is misleading to compare 1.000 and 0.9 without pointing out the difference in denominators.
    \item The faithfulness scores for Llama rag\_docs\_only and full\_agent are 0.755 and 0.750, respectively; Gemini scores 0.835 and 0.900.
\end{itemize}

\subsection{Latency}

Table~\ref{tab:latency} reports the mean and median response latency for each model configuration.

\begin{table}[htbp]
\caption{Mean and Median Latency by Configuration}
\label{tab:latency}
\centering
\resizebox{\columnwidth}{!}{%
\begin{tabular}{llcc}
\toprule
\textbf{Model} & \textbf{Configuration} & \textbf{Mean (ms)} & \textbf{Median (ms)} \\
\midrule
Gemini 3.5 FL & plain\_llm      & 1,332 & 912  \\
              & rag\_docs\_only  & 5,910 & 4,094\\
              & full\_agent     & 6,415 & 5,998\\
\midrule
Llama 3.2     & plain\_llm      & 6,300 & 5,727 \\
              & rag\_docs\_only  & 18,028& 17,918\\
              & full\_agent     & 19,496& 19,609\\
\bottomrule
\end{tabular}%
}
\end{table}

Llama is slower than Gemini in all of three scenarios. The difference should be taken into account in addition to the inference environments: In Gemini, it runs on a cloud-based API, whilst in Llama it is installed locally using Ollama. We know what software they're running and how much RAM and disk space they have, but the logs do not include any of the hardware details or CPU or GPU utilisation, so we can't determine what the cause of the latency difference is. In both models, 300 attempts to complete the benchmarks were made and were successfull. The Q98 Llama \texttt{full\_agent} output, for instance, started with an empty \texttt{\{ \}} body and then had a valid text following.

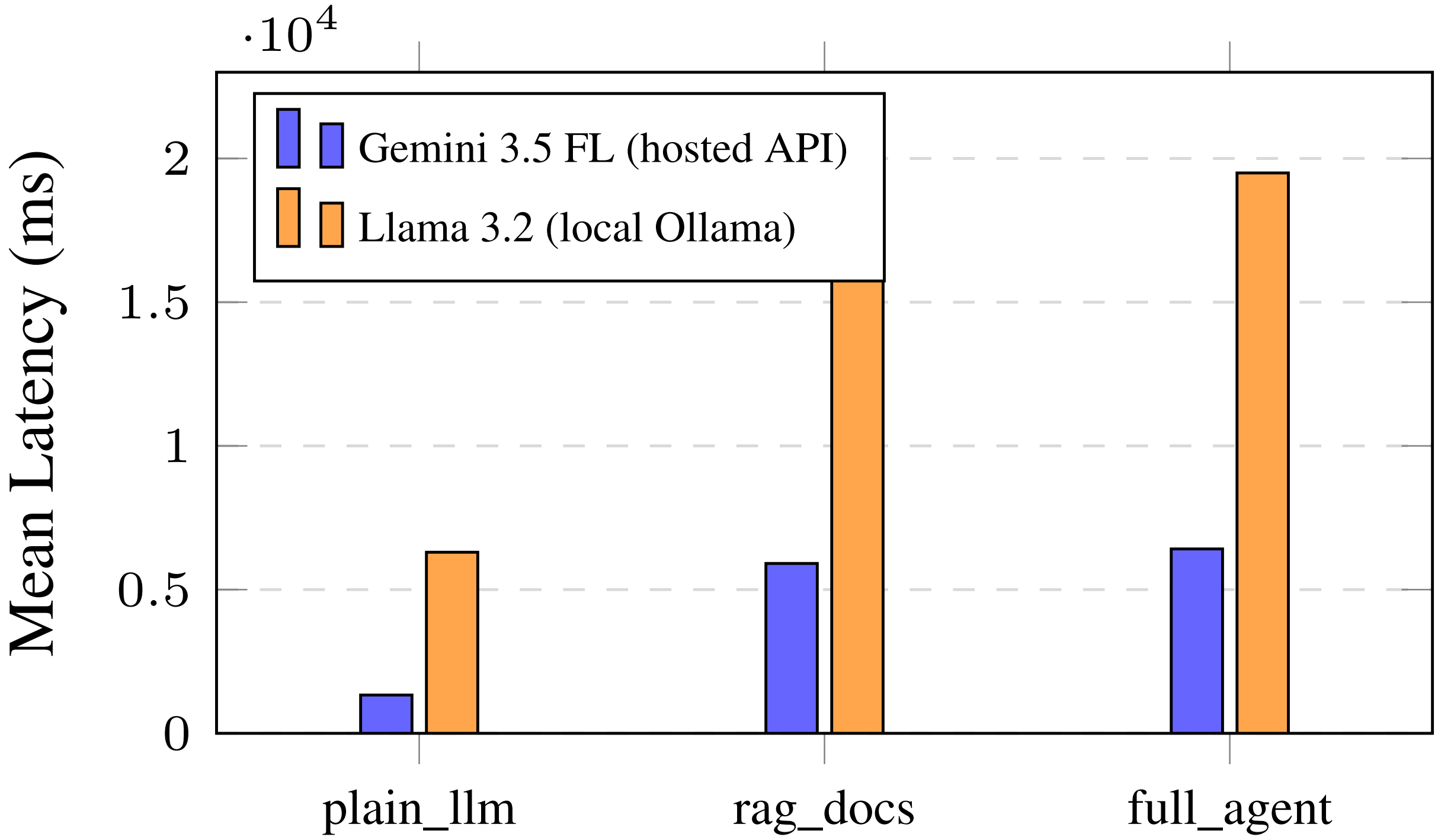
\begin{figure}[htbp]
\centering
\resizebox{0.98\columnwidth}{!}{%
\begin{tikzpicture}
\begin{axis}[
    ybar,
    bar width=0.25cm,
    width=7.5cm,
    height=4.8cm,
    ylabel={Mean Latency (ms)},
    ylabel style={font=\footnotesize},
    ymin=0, ymax=23000,
    ytick={0,5000,10000,15000,20000},
    yticklabel style={font=\scriptsize},
    xtick=data,
    xticklabels={plain\_llm, rag\_docs, full\_agent},
    xticklabel style={font=\scriptsize},
    legend style={at={(0.03,0.97)}, anchor=north west, font=\tiny, legend cell align=left},
    enlarge x limits=0.25,
    ymajorgrids=true,
    grid style={dashed, gray!30},
]
\addplot[fill=blue!60] coordinates {(1,1332)(2,5910)(3,6415)};
\addplot[fill=orange!70] coordinates {(1,6300)(2,18028)(3,19496)};
\legend{Gemini 3.5 FL (hosted API), Llama 3.2 (local Ollama)}
\end{axis}
\end{tikzpicture}%
}
\caption{Mean latency per question by model and configuration. Gemini uses a hosted cloud API; Llama uses local Ollama inference. Differences reflect inference environments (cloud API vs.\ local edge), not model execution speed alone.}
\label{fig:latency}
\end{figure}

Figure~\ref{fig:latency} represents the mean latency for each configuration.

\section{Discussion}
\label{sec:limitations}
\label{sec:discussion}

\subsection{Multi-Surface Tool Access Substantially Improves Gemini Performance}

The results for multi source questions Q53--Q84 ($n=32$) show a clear difference between the three Gemini configurations. The full-agent setup achieves 84.38\% correctness, compared with 40.63\% for document-only RAG. This corresponds to a gain of 43.75~pp, whereas the plain LLM does not answer any of these questions correctly (0.00\%). One reason for this difference is that several questions require information from more than one part of the workspace.

Q53 is a useful example. The Release Plan provides the release candidate date, but the feature owner's due date is recorded separately in the task tracker. Document-only RAG can retrieve the first detail but misses the task deadline, earning 0.5. The full agent, on the other hand, accesses both sources through \texttt{get\_tasks\_tool} and \texttt{search\_workspace\_tool}. It therefore retrieves both required details and receives 1.0.

\subsection{Observed Llama 3.2 Behavior Under Tool Access}

Llama~3.2 (3.2B, running locally through Ollama) shows a much smaller improvement from tool access. Under rubric v2.2-deterministic-audited, its gain over document-only RAG on the held-out set is 1.88~pp. For comparison, Gemini achieves a gain of 26.25~pp under the same evaluation.

These results alone do not explain why the two models perform differently. The comparison involves two particular model configurations, rather than an experiment designed to isolate the effects of model size or architecture. There are also differences in their execution setups that may have affected the outcome:

\begin{itemize}
    \item \textbf{Model configuration:} Gemini~3.5~Flash~Lite was accessed through its hosted API, while Llama~3.2 (3.2B) was run locally using Ollama. Since the two models were evaluated in different configurations, the experiment cannot isolate the effect of model scale.

    \item \textbf{Inference environment:} The inference environments were also different. Gemini used a cloud API, whereas Llama ran locally through Ollama. The GPU specifications of the local machine were not recorded, which prevents us from assessing the possible effect of hardware on the results.

    \item \textbf{Context window:} The Llama configuration had an 8,192-token limit. The difference in available context capacity may have influenced the results.

    \item \textbf{Tool-use capabilities:} The models may differ in how they were trained to select and interact with external tools. These differences could affect their ability to identify and use the appropriate workspace tools, but the present experiment does not examine this factor independently.
\end{itemize}

For these reasons, the observed gap should not be attributed exclusively to model size, architecture, or tool-use training. Separating the effects of these factors would require further experiments with controlled evaluation conditions.

\subsection{Task Retrieval Failure on Q37}

Q37 asks who is responsible for the document-export feature. In the \texttt{rag\_docs\_only} configuration, Llama retrieves Priya's ownership information from the Release Plan and receives 1.0. The full-agent configuration fails on the same question. After invoking \texttt{get\_tasks\_tool}, it reports that ``The document export task does not exist in the workspace.'' This is incorrect because task T01 (``Document export'', owner Priya) is present in the database.

The response consequently receives 0.0, contributing to the -2.50~pp regression across the 40 questions. This case shows that invoking a tool does not necessarily lead to successful retrieval or answer synthesis. However, the failure on Q37 is a single observation and cannot, by itself, establish a broader pattern of tool distraction.

\subsection{Configuration-Specific Hallucination Patterns}

The raw responses also reveal cases where the models provide details that cannot be supported by the available workspace information. Some examples are particularly noticeable:

\begin{itemize}
    \item \textbf{Q85:} Llama \texttt{rag\_docs\_only} introduces November 16, 2026, as the mobile app release date. Its response states, ``The target production launch date is 16 November 2026,'' applying a date incorrectly to a question about a mobile app. The full-agent configuration avoids this unsupported answer and responds, ``Unfortunately, I don't have information on upcoming releases of a specific Orbit mobile app version.''

    \item \textbf{Q93:} The name ``Omar'' is introduced as the on-call engineer in the Llama \texttt{full\_agent} response, despite the absence of supporting information. Llama \texttt{rag\_docs\_only} correctly identifies that the documents do not contain this information. The plain LLM also abstains, although this response is not counted as correct under rubric v2.2.

    \item \textbf{Q99:} In the full-agent configuration, Llama invents an approval record: ``The final press release wording was approved by Omar on 15 November 2026, at 10:00 AM UTC.'' The plain LLM abstains, but its response is not credited as correct under rubric v2.2. The document-only configuration produces a different unsupported answer by referring to a ``Release Manager'' without identifying a name or date. This is classified as a mixed response and retains a score of 0.0.

    \item \textbf{Q91 and Q92:} Both Llama \texttt{rag\_docs\_only} and \texttt{full\_agent} provide specific dates for security incidents and EU regulation timelines that are not documented in the workspace. These cases are genuine hallucinations and are separate from the natural-language extractions corrected under v2.2-deterministic-audited.
\end{itemize}

\subsection{Gemini Scoring Correction}

The v2.2 correction affected the reported Gemini full-agent scores. On the complete dataset, the score changed from 72.50\% to 74.50\%. For the held-out cohort, it increased from 83.75\% to 86.25\%. The corrected figures are used throughout this paper, while the original v2.1 results have been retained in \url{scored_phase3_benchmark_gemini_35_lite_20261025_165427.json}.


\section{Conclusion}
\label{sec:conclusion}

This study evaluated Orbit AI, a tool-augmented RAG system designed to answer questions by retrieving information from different enterprise workspace resources. Its performance was assessed using a benchmark of 100 questions and a deterministic scoring procedure.

\textbf{Primary findings from the held-out cohort (Q21--Q100):}

\begin{itemize}
    \item In the full-agent configuration, Gemini 3.5 Flash Lite achieved 86.25\% overall correctness and 84.38\% on multi-source questions. On the same multi-source questions, document-only RAG achieved 40.63\%. The resulting difference of 43.75 percentage points indicates the value of accessing multiple workspace resources when answering questions that combine information from documents, task records, and chat messages.

    \item Llama 3.2 (3.2B), evaluated locally, recorded 52.50\% overall correctness on the held-out cohort. Its improvement over document-only RAG was considerably smaller, at 1.88 percentage points. This difference should be interpreted with care, since the two model setups also used different inference environments and context-window limits.

    \item All 300 benchmark attempts completed successfully for both models, giving an execution success rate of 100\%. Nevertheless, successful execution alone does not establish the correctness of a generated answer.
\end{itemize}

\textbf{Findings related to the evaluation method:}

\begin{itemize}
    \item Under rubric v2.2-deterministic-audited, Llama~3.2 achieves 5.50\% overall correctness. One limitation of the scoring procedure is that its deterministic refusal detector may fail to identify some valid abstentions expressed in natural language.

    \item Two Gemini responses from the held-out cohort (Q93 and Q99) were rescored following an update to the scoring rules. Consequently, the held-out cohort was not entirely independent of the scoring process.
\end{itemize}

\textbf{Future work} will need to address both the scale of the evaluation and the limitations of the current methodology. Testing on larger and more diverse workspaces, together with repeated runs of each configuration, would make it possible to examine performance variability and report confidence intervals. Human evaluation could provide additional insight into borderline responses, while separate tool-level experiments could establish the individual contributions of document search, task lookup, and chat retrieval. The influence of the Q53 example in the system prompt also deserves investigation. In addition, evaluating Llama with GPU-accelerated inference could provide further insight into its performance under a different inference environment.

Throughout this study, the reported findings have been kept consistent with the recorded experimental results. Claims that could not be verified were excluded, and the scoring changes introduced after the initial evaluation have been explicitly documented.

\subsection{Future Work and Scope}
\label{sec:future-work}

The present evaluation also raises several directions for further investigation:

\begin{enumerate}
    \item \textbf{Larger workspace scale:} The current evaluation uses a relatively small synthetic workspace. A useful next step would be to test Orbit AI on larger project environments containing thousands of documents, tasks, and chat messages. Such experiments could reveal how increasing workspace size affects retrieval precision, token consumption, and multi-step reasoning.

    \item \textbf{Multi-chunk document retrieval:} In Phase 3, each document fits within a single chunk of approximately 1,000 words. This does not fully test retrieval across document boundaries or multiple chunks. Using longer documents split into several chunks would allow future experiments to examine cross-chunk retrieval, the combination of information from separate chunks, and retrieval precision at a larger scale.

    \item \textbf{Uniform question cohorts:} Comparisons between RAG and agent configurations would benefit from using an identical set of questions throughout the evaluation. This is particularly relevant to Q01--Q20, where some answers were no longer available in the current workspace version, resulting in cohort misalignment. The held-out cohort provides a basis for evaluating the configurations more uniformly.

    \item \textbf{Improved refusal scoring:} The deterministic v2.2 rubric could be expanded to cover a broader range of natural-language refusal expressions. This would reduce the likelihood of valid abstentions being marked incorrect simply because their wording does not match the existing refusal signal list.

    \item \textbf{Independent held-out evaluation:} Scoring rules should ideally be finalized before examining responses from the held-out cohort. For example, the Gemini responses to Q93 and Q99 could be assessed using rules established in advance. This would reduce the influence of post hoc scoring changes and preserve greater independence between rubric development and held-out evaluation.

    \item \textbf{System prompt evaluation:} The role of system prompt examples in model performance remains worth examining. One specific experiment could remove the example concerning ``Priya's document export task due date'' and compare the resulting responses to Q53. This would help determine whether including that example influences performance on the question.

    \item \textbf{Repeated experimental runs:} Each configuration was evaluated only once in this study. Running every configuration multiple times would provide a better picture of performance variability. The resulting measurements could also support confidence interval estimation and statistical significance testing.

    \item \textbf{Human evaluation:} Automated scoring could be complemented by blinded human assessment. This would be particularly useful for responses involving partial refusals or answers containing valid information alongside unsupported details. Comparing human judgments with automated scores could also reveal disagreements in the evaluation process.

    \item \textbf{Controlled inference environments:} A more comparable inference setup for Gemini and Llama would help investigate how much of the observed performance and latency difference comes from the execution environment rather than the underlying model architectures.

    \item \textbf{Handling empty or malformed responses:} The evaluation procedure also needs a consistent approach to responses that are empty or incorrectly formatted. One example is the Llama Q98 response in the \texttt{full\_agent} configuration, which begins with ``\{\}''. Further investigation could establish how such outputs should be processed and scored.

    \item \textbf{Improved scoring methods:} The current keyword- and regex-based scoring procedure could be extended to recognize semantically equivalent expressions. It should also distinguish genuine contradictions from values that are mentioned only to provide additional context. This could improve the consistency of automated evaluation.

    \item \textbf{Broader evaluation scope:} Finally, the evaluation could be extended to additional benchmarks, models, configurations, and scoring approaches. Testing across these different settings would help determine whether the observations from the current synthetic benchmark also hold in other environments.
\end{enumerate}
\section*{Code Availability}

The source code and implementation of Orbit AI are publicly available at:

\url{https://github.com/avadhesh11/Design-Project}.

The questionarrie is available at 
\href{https://github.com/avadhesh11/Design-Project/blob/main/backend/evaluation/questions_phase3_100.json}{Click Here}

\bibliographystyle{unsrtnat}
\bibliography{references}

\end{document}